\documentstyle[12pt,epsf]{article}

\begin{document}


\begin{center}

 {\Large \bf
\vskip 7cm
\mbox{Exclusive Double Diffractive Events:}
\mbox{ Menu for LHC.}
}
\vskip 1cm

\vskip 3cm

\mbox{Petrov~V.A. and Ryutin~R.A.}

\mbox{{\small Institute for High Energy Physics}}

\mbox{{\small{\it 142 281} Protvino, Russia}}

 \vskip 1.75cm
{\bf
\mbox{Abstract}}
  \vskip 0.3cm

\newlength{\qqq}
\settowidth{\qqq}{In the framework of the operator product  expansion, the quark mass dependence of}
\hfill
\noindent
\begin{minipage}{\qqq}
Exclusive double diffractive events (EDDE) are considered in the framework of 
the Regge-eikonal approach and perturbative calculations for "hard" subprocesses. Total 
and differential cross-sections for processes $p+p\to p+X+p$ are calculated. 

\end{minipage}
\end{center}


\begin{center}
\vskip 0.5cm
{\bf
\mbox{Keywords}}
\vskip 0.3cm

\settowidth{\qqq}{In the framework of the operator product  expansion, the quark mass dependence of}
\hfill
\noindent
\begin{minipage}{\qqq}
Exclusive Double Diffractive Events -- Pomeron -- Regge-Eikonal model -- Higgs -- Radion -- Jets
\end{minipage}

\end{center}

\setcounter{page}{1}
\newpage


\section{Introduction}

LHC collaborations aimed at working in low and high $p_T$
regimes related to typical undulatory (diffractive) and corpuscular (point-like) 
behaviours of the corresponding cross-sections may offer a very exciting possibility to
observe an interplay of both regimes~\cite{i1}. In theory the "hard part" can be
(hopefully) treated with perturbative methods whilst the "soft" one is definitely
nonperturbative.

 Below we give several examples of such an interplay: exclusive particle production by diffractively
scattered protons, i.e. the processes $p+p\to p+X+p$, where $+$ means also a rapidity 
gap and X represents a particle or a system of particles consisting of or strongly
coupled to the two-gluon state.

 This process is related to the dominant amplitude of exclusive two-gluon 
production. Driving mechanism of the diffractive processes is the Pomeron. Data on the total
cross-sections demands unambiguosly for the Pomeron with larger-than-one intercept, thereof
the need in "unitarisation". 

As will be seen below, EDDE gives us unique experimental possibilities for particle searches
and investigations of diffraction itself. This is due to several advantages of the process: a) clear
signature of the process; b) possibility to use "missing mass method", that improve the mass 
resolution; c) background is strongly suppressed; d) spin-parity analysis of the central system can be 
done; e) interesting measurements concerning the interplay between "soft" and "hard" scales are possible.

\section{Calculations}

In Figs.~1,2 we illustrate in detail the process
$p+p\to p+X+p$. Off-shell proton-gluon amplitudes in Fig.~1 are treated by the method developed
in Ref.~\cite{i2}, which is based on the extension of Regge-eikonal approach, and 
succesfully used for the description of the HERA data~\cite{i3}.   

 The amplitude of the process $p+p\to p+X+p$ can be obtained in the following way (see~Figs.~1,2). The first
step is to calculate the "bare" amplitude $T_X$, which is depicted in Fig.~1. The "hard" 
part is the usual gluon-gluon fusion process 
calculated by perturbative methods in the Standard Model or its extensions. "Soft" amplitudes $T_{1,2}$ are obtained 
in the Regge-eikonal approach. The second step is the unitarization procedure, that takes into account
initial and final state interactions (see Fig.~2).
  
 We use the following kinematics, which corresponds to the double Regge limit. It is convenient
to use light-cone components $(+,-;\bot)$. The components of momenta of the hadrons in Fig.~1 are
\begin{eqnarray}
\label{momenta}
p_1&=&\left( \sqrt{\frac{s}{2}},\frac{m^2}{\sqrt{2s}}, {\bf 0}\right)\\
p_2&=&\left( \frac{m^2}{\sqrt{2s}},\sqrt{\frac{s}{2}}, {\bf 0}\right)\nonumber\\
p_1^{\prime}&=&\left( (1-\xi_1)\sqrt{\frac{s}{2}},\frac{{\bf\Delta}_1^2+m^2}{(1-\xi_1)\sqrt{2s}}, -{\bf\Delta}_1\right)\nonumber\\
p_2^{\prime}&=&\left( \frac{{\bf\Delta}_2^2+m^2}{(1-\xi_2)\sqrt{2s}},(1-\xi_2)\sqrt{\frac{s}{2}}, -{\bf\Delta}_2\right)\nonumber\\
q&=&\left( q_+, q_-, {\bf q}\right)\;{,}\nonumber\\
q_1&=&q+p_1-p_1^{\prime}=q+\Delta_1\;{,}\nonumber\\
q_2&=&-q+p_2-p_2^{\prime}=-q+\Delta_2\;{,}\nonumber
\end{eqnarray}
$\xi_{1,2}$ are fractions of protons momenta carried by gluons. For two-dimensional 
transverse vectors we use boldface type. From the above notations we can
obtain the relations:
\begin{eqnarray}
\label{notations}
t_{1,2}\;\;=\;\;\Delta_{1,2}^2 &\simeq& -\frac{{\bf\Delta}_{1,2}^2(1+\xi_{1,2})+\xi_{1,2}^2m^2}{1-\xi_{1,2}}\;\;\simeq\\
&\simeq& -{\bf\Delta}_{1,2}^2\;{,}\;\;\xi_{1,2}\to 0\nonumber\\
\cos\phi_0&=&\frac{{\bf\Delta}_1{\bf\Delta}_2}{|{\bf\Delta}_1||{\bf\Delta}_2|}\nonumber\\
M_X^2&\simeq& \xi_1\xi_2s+t_1+t_2-2\sqrt{t_1t_2}\cos\phi_0\nonumber\\
(p_1+q)^2&\simeq&m^2+q^2+\sqrt{2s}q_-=s_1\nonumber\\
(p_2-q)^2&\simeq&m^2+q^2-\sqrt{2s}q_+=s_2\; .\nonumber
\end{eqnarray}
Physical region of diffractive events with two rapidity gaps is defined by the following  
kinematical cuts:
\begin{equation}
\label{tlimits}
0.01\; GeV^2\le |t_{1,2}|\le 1\; GeV^2\;{,} 
\end{equation}
\begin{equation}
\label{xlimits}
\xi_{min}\simeq\frac{M_X^2}{s \xi_{max}}\le \xi_{1,2}\le \xi_{max}=0.1\;,
\end{equation} 
\begin{eqnarray}
\label{klimits}
\left(\sqrt{-t_1}-\sqrt{-t_2}\right)^2\le\!&\kappa&\!\le\left(\sqrt{-t_1}+\sqrt{-t_2}\right)^2\\
\kappa=\xi_1\xi_2s\!\!&-&\!\! M_X^2\ll M_X^2\nonumber
\end{eqnarray}
We can write the relations in terms of $y_{1,2}$ and $y_X$ (rapidities of hadrons and the system X correspondingly). For instance:
\begin{eqnarray}
\label{raplimits}
&\xi_{1,2}&\!\!\simeq\frac{M_X}{\sqrt{s}}e^{\pm y_X}\; ,\\
&|y_X|&\!\!\le y_0=\ln\left(\frac{\sqrt{s}\xi_{max}}{M_X}\right)\; ,\nonumber\\
y_0\simeq 2.5\!&\mbox{\rm for}&\!\sqrt{s}=14\; TeV\;{,}\; M_X=O(100\; GeV)\; , \nonumber\\
&|y_{1,2}|&\!\!=\frac{1}{2}\ln\frac{(1-\xi_{1,2})^2s}{m^2-t_{1,2}}\ge 9\nonumber
\end{eqnarray}
 
 In standard terms the amplitude corresponds to the so-called nonfactorized scheme~\cite{Collins}. The 
contribution of the diagram depicted in Fig.~2
is obtained by integrating over all internal loop momenta. It was 
shown in~\cite{Collins}, that the leading contribution arises from the region of the
integration, where momentum $q$ is "Glauber-like", i.e. of the order 
$(\mbox{k}_+m^2/\sqrt{s},\mbox{k}_-m^2/\sqrt{s},\mbox{\bf k}m)$, where k's are of the 
order $1$. The detailed consideration of the loop integral like
\begin{equation}
\int\frac{d^4 q}{(2\pi)^4}\frac{f(q,p_1,p_2,\Delta_1,\Delta_2)}{(q^2+i 0) (q_1^2+i 0) (q_2^2+i 0)}
\end{equation}
shows that the main contribution comes from the poles at
\begin{eqnarray}
&q_1^2&\!\!\!=\sqrt{2s}\xi_1q_{-}-{\mbox{\bf q}_1}^2 = 0\; ,\nonumber\\ 
&q_2^2&\!\!\!=-\sqrt{2s}\xi_2q_{+}-{\mbox{\bf q}_2}^2 = 0\; .\nonumber
\end{eqnarray}
In this case
\begin{equation}
\label{qvalue}
q=\left(-\frac{{\mbox{\bf q}_2}^2}{\xi_2\sqrt{2s}},\; \frac{{\mbox{\bf q}_1}^2}{\xi_1\sqrt{2s}}, \mbox{\bf q}
\right)\;{,}
\end{equation}
where 
\begin{eqnarray}
&{\mbox{\bf q}_1}^2&\!\!\!\!={\mbox{\bf q}}^2+{{\bf\Delta}_1}^2+2|\mbox{\bf q}||{\bf\Delta}_1|\cos(\phi+\frac{\phi_0}{2})\;{,}\nonumber\\
&{\mbox{\bf q}_2}^2&\!\!\!\!={\mbox{\bf q}}^2+{{\bf\Delta}_2}^2-2|\mbox{\bf q}||{\bf\Delta}_2|\cos(\phi-\frac{\phi_0}{2})\nonumber
\end{eqnarray}

 Taking the general form for $T$-amplitudes that satisfy identities
\begin{equation} 
\label{wardid}
q^{\alpha}T^D_{\mu\alpha}=0,\; q_i^{\mu}T^D_{\mu\alpha}=0\;, 
\end{equation}
and neglecting terms of the order
$o(\xi_i)$, the following expression is found at $|t_i|\le 1$~GeV$^2$:
\begin{equation}
\label{Tamp}
T^{D}_{\mu\alpha}(p,\; q,\; q_i)=\left( G_{\mu\alpha} - \frac{P^{q_i}_{\mu}P^{q}_{\alpha}}{P^{q_i}P^q} 
\right) T^{D}_{gp\to gp}(s_i,t_i,qq_i)
\end{equation}
$$
\noindent G_{\mu\alpha}=g_{\mu\nu}-
\frac{q_{i,\mu}q_{\alpha}}{qq_i}\; ,
$$
$$
\noindent P^{q_i}_{\mu}=p_{\mu}-\frac{pq_i}{qq_i}q_{\mu}\; ,
$$
$$
\noindent P^{q}_{\alpha}=p_{\alpha}-\frac{pq}{qq_i}q_{i,\alpha}\; .
$$
For $T^{D}_{gp\to gp}$ we use the
Regge-eikonal approach~\cite{i2,3Pomerons}. At small $t_i$ it takes the form
of the Born approximation, i.e. Regge factor: 
\begin{equation}
\label{REapproach}
T^D_{gp\to gp}(s_i,t_i,qq_i)= c_{gp}
\left(e^{-i\frac{\pi}{2}}
\frac{s_i-qq_i-m^2}{s_0-qq_i-m^2}
\right)^{\alpha_P(t_i)}
e^{b_0t_i}\;,
\end{equation}
$$
b_0=\frac{1}{4}(\frac{r^2_{pp}}{2}+r^2_{gp})\;,
$$
where $\alpha_P(0)=1.203$, $\alpha_P^{\prime}(0)=0.094$~GeV$^{-2}$, $r^2_{pp}=2.477$~GeV$^{-2}$ are fixed 
parameters for the "hard" Pomeron~\cite{3Pomerons}, which have been obtained from the global fit to the data 
on diffractive $pp(p\bar{p})$ scattering. Parameters 
$c_{gp}\simeq 3.5$, $r^2_{gp}=2.54$~GeV$^{-2}$ are defined from fitting 
the HERA data on elastic $J/\Psi$ production~\cite{HERA}, which will be published elsewhere. The upper bound for the 
constant $c^{up}_{gp}\simeq 2.3(3.3)$ can be also estimated from
the exclusive double diffractive di-jet production at Tevatron (see~(\ref{dijetres})), if we take CDF cuts and the
upper limit for the exclusive total di-jet cross-section~\cite{CDF}. The effective value $c_{gp}=2.3$ corresponds 
to the case, when the Sudakov suppression factor is absorbed into the constant, and $c_{gp}=3.3$ is obtained 
when taking into account this factor explicitely.

 The full "bare" amplitude looks as follows:
\begin{equation}
\label{Tpp_pXp}
T_{pp\to pXp}\simeq
\int\frac{d^4 q}{(2\pi)^4}\frac{8 F^{\mu\nu}(q_1,q_2)T^{D}_{\mu\alpha}(p_1,\; q,\; q_1)T^{D}_{\nu\alpha}(p_2,\; -q,\; q_2)}{(q^2+i 0) (q_1^2+i 0) (q_2^2+i 0)}\; ,
\end{equation}
where 
$$
F^{\mu\nu}(q_1,q_2)=(g^{\mu\nu}-\frac{q_2^{\mu}q_1^{\nu}}{M_X^2})F_{gg\to X}\;.
$$ 
Factor $8$ arises from the colour index contraction. Let $l^2=-q^2\simeq{\mbox{\bf q}}^2$, $y_X=<y_X>=0$ and
contract all the tensor indices, then the integral~(\ref{Tpp_pXp})
takes the form

\begin{equation}
\label{Tpp_pXp2}
T_{pp\to pXp}\simeq
c_{gp}^2 e^{b(t_1+t_2)}\frac{\pi}{(2\pi)^2}
\left(-\frac{s}{M_X^2}\right)^{\alpha_P(0)}\cdot 8 F_{gg\to X}\cdot I\; ,
\end{equation}
\begin{equation}
b= \alpha^{\prime}_P(0) \ln\left(\frac{\sqrt{s}}{M_X}\right)+b_0\; ,
\end{equation}
\begin{equation}
\label{II}
I\simeq \int_{0}^{M_X^2}\frac{dl^2}{l^4} 
\left(\frac{l^2}{s_0-m^2+l^2/2}\right)^{2\alpha_P(0)}\;,
\end{equation}
where $s_0-m^2\simeq 1$~GeV$^2$ is the scale parameter of the model that is used in the global fitting of 
the data on $pp(p\bar{p})$ scattering for on-shell amplitudes~\cite{3Pomerons}. It remains fixed in the present 
calculations. It worth nothing that the "rescattering corrections" for the off-shell
gluon-proton amplitudes, $T^D$, are small (in accordance with a general analysis in Ref.~\cite{i2}). Contrary
to this, the "outer" corrections (see Eq.(\ref{ucorr})) are significant.

If we take into account the emission of virtual "soft" gluons, while prohibiting the real ones, that 
could fill rapidity gaps, it results in
the Sudakov-like suppression~\cite{Khoze}:

\begin{eqnarray}
\label{sudakov}
F_s(l^2)=exp\left[-\frac{3}{2\pi}\int\limits_{l^2}^{{M_X}^2/4}\frac{d{p_T}^2}{{p_T}^2}\alpha_s({p_T}^2)\ln\left(
\frac{{M_X}^2}{4{p_T}^2}\right)\right]\; ,
\end{eqnarray}
and in the new value of the integral~(\ref{II}):
\begin{equation}
\label{IInew}
I_s\simeq \int_{0}^{M_X^2}\frac{dl^2}{l^4} F_s(l^2) 
\left(\frac{l^2}{s_0-m^2+l^2/2}\right)^{2\alpha_P(0)}\; .
\end{equation}
In this case the total cross-section becomes smaller, than without the factor $F_s$. It plays
significant role for large $M_X$.

Unitarity corrections can be estimated from the elastic $pp$ 
scattering by the method depicted in Fig.2, where

\begin{eqnarray}
\label{ucorr}
T_X&=&T_{pp\to pXp}\;,\\
V(s\;,\;\mbox{\bf q}_T)&=&4s(2\pi)^2\delta^2(\mbox{\bf q}_T)
+4s\int d^2\mbox{\bf b}e^{i\mbox{\bf q}_T \mbox{\bf b}}
\left[e^{i\delta_{pp\to pp}}-1\right]\;,\nonumber\\
T^{Unit.}_X(p_1\;,\; p_2\;,\;\Delta_1\;,\;\Delta_2)
&=&\frac{1}{16ss^{\prime}}\int
\frac{d^2\mbox{\bf q}_T}{(2\pi)^2}\frac{d^2\mbox{\bf q}^{\prime}_T}{(2\pi)^2}
V(s\;,\;\mbox{\bf q}_T)\;
\cdot\;T_X( p_1-q_T, p_2+q_T,\Delta_{1T},
\Delta_{2T})\;\cdot\nonumber\\
&\cdot&\; V(s^{\prime}\;,\;\mbox{\bf q}^{\prime}_T)\;,\nonumber\\
\Delta_{1T}&=&\Delta_{1}-q_T-q^{\prime}_T\;,\nonumber\\
\Delta_{2T}&=&\Delta_{2}+q_T+q^{\prime}_T\;,\nonumber
\end{eqnarray}
and $\delta_{pp\to pp}$ can be found in Ref.~\cite{3Pomerons}. These "outer" unitarity corrections
reduce the integrated cross-section 
by the factor about $14$ for the given kinematical cuts and lead to the changes in the $\phi_0$-dependence. 

\section{Results for resonance production} 

 We have the following expression for the differential cross-section in case of one particle production:
\begin{eqnarray}
\label{dsigttxx}
\frac{d\sigma}{dt_1dt_2d\xi_1d\xi_2}\!\!\!\!\!\!\!\!\!\!\!\!& &=\frac{\pi |T^{Unit.}_{pp\to pXp}|^2}{8s(2\pi)^5\sqrt{-\lambda}}\\
&\lambda&=\kappa^2+2(t_1+t_2)\kappa+(t_1-t_2)^2\le 0\nonumber
\end{eqnarray}
 
 By partial integrating~(\ref{dsigttxx}) we obtain $t$ and $\xi$ distributions. The first 
result of our calculations is depicted 
in the Fig.~3. The antishrinkage of the 
diffraction peak with
increasing mass $M_X$ is the direct consequence of the existence of the additional
hard scale $M_X$, which makes the interaction radius smaller. The $\xi$ distribution
is shown in Fig.4.

We can use the following replacement to obtain the cross-section for the EDD process $p+p\to p+X+p$:
\begin{equation}
\label{FggtoX}
|F_{gg\to X}|^2\to 4\pi M_X\Gamma(X\to gg)\;.
\end{equation}

It is possible to simplify calculations after reduction of~(\ref{dsigttxx}) to
\begin{equation}
\label{reddsig}
\frac{d\sigma}{dt_1dt_2d\phi_0}\simeq\frac{\pi |T^{Unit.}_{pp\to pXp}|_{y_X=0}^2}{8s^2(2\pi)^5}\Delta y_X\; ,
\end{equation}
where $\Delta y_X=2 y_0$, $\phi_0$ is the azimuthal angle between outgoing protons.

\section{Standard model Higgs boson production}

For the Standard Model Higgs boson~\cite{SMHiggs}
\begin{eqnarray}
\label{ggHvertex}
\!\!F^0_{gg\to H}&\!\!=\!\!& M_H^2\frac{\alpha_s}{2\pi}\sqrt{\frac{G_F}{\sqrt{2}}}f(\eta)\;,\\
\!\!f(\eta)&\!\!=\!\!&\frac{1}{\eta}\left\{ 1+\frac{1}{2}\left( 1-\frac{1}{\eta} \right) \left[ 
Li_2\left( \frac{2}{1-\sqrt{1-\frac{1}{\eta}}-i0} \right)+
Li_2\left( \frac{2}{1+\sqrt{1-\frac{1}{\eta}}+i0} \right)
\right]
\right\}\; ,\nonumber\\
|F_{gg\to H}|^2&\to&1.5 |F^0_{gg\to H}|^2\; ,
\end{eqnarray}
where $\eta=M_H^2/4 m_t^2$, $G_F$ is the Fermi constant, $m_t$ is the top quark mass. NLO K-factor 1.5 for the $gg\to H$ process 
is included to the final answer.

Numerical results are the following~\cite{EDDEH}

\vspace*{0.4cm}
\begin{tabular}{|c|c|c|c|c|c|}
\hline
          &                 &\multicolumn{4}{|c|}{$\sigma_{p+p\to p+H+p}$ (fb)}\\
\cline{3-6} 		  
$c_{gp}$  &  $M_H$ (GeV)    &\multicolumn{2}{|c|}{LHC}      &\multicolumn{2}{|c|}{TeVatron}\\
\cline{3-6}
          &                 & no Sud. suppr.& Sud. suppr.     & no Sud. suppr.   & Sud. suppr.\\
\hline 
   3.5  &  $100\to 500$   &  $110\to 57$  & $4.6\to 0.14$  & $12\to 0.4$   & $0.5\to 0.001$ \\
\hline 
   2.3(3.3) &  $100\to 500$   &  $20\to 11$ & $\mbox{\bf 3.6}\to \mbox{\bf 0.11}$  & $2.2\to 0.08$     & $0.4\to 0.0009$  \\
\hline 
\end{tabular}
\vspace*{0.4cm}

We consider four different cases only for Higgs boson as illustration of the Sudakov 
suppressing action. In other examples we 
take $c_{gp}=3.3$ with Sudakov-like 
suppression according to the CDF data estimations to obtain the lower value of cross-sections at LHC and
Tevatron. For this case our result is quite close to the one of~\cite{Khoze}, where the value of the total cross-section is 
about $3$~fb. In both cases the most important suppressing in the mass region $M_H>100$~GeV is due to
(perturbative) Sudakov factors, while the nonperturbative (absorbtive) factors play relatively
minor role. 

 Results of other authors were considered in details in~\cite{Khozemyths}. Here we refer to the 
highest cross-section $2$~pb for $M_H=400$~GeV at LHC energies that was 
obtained in Ref.~\cite{Cudell}. A nonfactorized
form of the amplitude and a "QCD inspired" model for $g p\to g p$ amplitudes were used, taking into account
the nonperturbative proton wave functions. Even if we multiply the result of Ref.~\cite{Cudell} by 
the suppressing factor, it will be larger than ours. This could serve as the indication of
the role of nonperturbative effects. Our model is based on the Regge-eikonal approach for
the amplitudes, which is primordially nonperturbative, normalized to the 
data from HERA on $\gamma p\to J/\Psi p$~\cite{HERA} and improved by the CDF data on the exclusive di-jet production~\cite{CDF}.  

 To estimate the signal to QCD background ratio 
for $b\bar{b}$ signal we use the standard expression for $gg\to b\bar{b}$ amplitude and assumptions~\cite{Khozebg1chi}-\cite{KhozeJz0}:
\begin{itemize}
\item possibility to separate final $b\bar{b}$ quark jets from gluon jets. If we cannot
do it, it will increase the background by two orders of magnitude under the $50\%$ efficiency.
\item suppression due to the absence of colour-octet $b\bar{b}$ final 
states 
\item suppression of light fermion pair production, when $J_{z,tot}=0$ (see also~\cite{Jz0lvanish1},\cite{Jz0lvanish2})
\item cut $E_T>50$~GeV ($\theta\ge 60^{\mbox{\small o}}$), since the cross-section of
EDD $b\bar{b}$ jet production strongly decreases with $E_T$ (see formulae (\ref{qqdijet})).    
\end{itemize}
The theoretical result of our numerical estimations is 
\begin{equation}
\label{sigbgex}
\frac{Signal(pp\to pHp\to pb\bar{b}p)}{QCD\;\;background}\ge 3.8 \frac{GeV}{\Delta M}\;{,} 
\end{equation}
where $\Delta M$ is the mass resolution of the detector and $M_H\simeq 115$~GeV, which can reach $0.01 M_H$ due
to application of the "missing mass method". Similar result was strictly obtained 
in~\cite{Khozebg1chi},\cite{Khozebg2full}. Under the above circumstances the total efficiency at the integrated luminocity $30$~fb$^{-1}$
is $\sim10\%$ and numerically estimated significance of the event is about $3\sigma$, which is close
to the one in $\gamma\gamma$ decay mode. 

\section{Heavy quarkonium production}

Results for $\chi_{c0,b0}$ EDD production were obtained recently
by some authors~\cite{Khozebg1chi},\cite{KhozeJz0chi},\cite{Chi_other} in different 
approaches. To obtain the total cross-sections in the model considered in the present 
paper let us substitute widths of these states into~(\ref{FggtoX}).

\begin{equation}
\Gamma(\chi_{b0}\to gg)\simeq\Gamma_0(\chi_{b0}\to gg) \left( 1+9.8\frac{\alpha_S}{\pi}\right)=550\; keV\; \mbox{(see \cite{Khozebg1chi},\cite{KhozeJz0chi} for details)},
\end{equation}
where the width is set to the lattice result $\Gamma_0(\chi_{b0}\to gg)=354\; keV$~\cite{latchi_b}. After
replacement we obtain for LHC and TeVatron: 

\begin{equation}
\sigma_{pp\to p+\chi_{b0}+p}\simeq 1.3\; nb\;, \sqrt{s}=14\; TeV\;,\;\mbox{((\ref{tlimits}),(\ref{xlimits}) cuts),}
\end{equation}

\begin{equation}
\sigma_{pp\to p+\chi_{b0}+p}\simeq 160\; pb\;, \sqrt{s}=1.8\; TeV\;,\; \mbox{(CDF cuts),}
\end{equation}

The same procedure can be done for $\chi_{c0}$. Taking the total width 
$\Gamma(\chi_{c0}\to gg)\simeq 14.9\; MeV$~\cite{chi_cwidthPDG} we obtain

\begin{equation}
\sigma_{pp\to p+\chi_{c0}+p}\simeq 4\;\mu b\;, \sqrt{s}=14\; TeV\;,\;\mbox{((\ref{tlimits}),(\ref{xlimits}) cuts),}
\end{equation}

\begin{equation}
\sigma_{pp\to p+\chi_{c0}+p}\simeq 600\;nb\;, \sqrt{s}=1.8\; TeV\;,\; \mbox{(CDF cuts).}
\end{equation} 

\section{Radion production.}

Now there is a great interest to the multidimensional properties of the space-time. One of the
models was proposed by Randall and Sundrum~\cite{RS1TH}. We have considered the case of one
compact extra-dimension. In this case we have additional scalar particle Radion, that
reflects the existence of an extra dimension and represents the field of the "distance" oscillations 
between the branes along the extra  
dimension. Since Radion has the same quantum numbers as the Higgs boson, they can mix~\cite{RS1TH2}.
After mixing we have two mass eigenstates, which could be observed experimentally. For the EDDE
the following replacements in (\ref{ggHvertex}) should be done:
\begin{eqnarray}
\label{hstar}
f(\eta)&\to& a_{34} f(\eta_{h^*})+ 7 \gamma\; b\;\; \mbox{for}\; h^*\;,\\
f(\eta)&\to& \gamma\left( a_{12} f(\eta_{r^*})+ 7 a\right)\;\; \mbox{for}\; r^*\;,
\end{eqnarray}
where $\eta_{h^*,r^*}=m_{h^*,r^*}^2/4 m_t^2$ and other parameters are obtained 
from formulae in the Appendix A of~\cite{RS1TH2}:

\begin{eqnarray}
\label{RS1pars}
&&\gamma=v/\Lambda_{\phi}\;, v=246\;GeV\;\mbox{is the Higgs VEV}\;,\; \Lambda_{\phi}\;\mbox{is the radion VEV}\;,\\
&&Z^2=1-6\xi\gamma^2(1+6\xi)\;,\;\tan 2\theta=12\xi\gamma Z\frac{1}{Z^2-36\xi^2\gamma^2-m^2_r/m^2_h}\;,\\
&&\nonumber\\
&&a=\cos\theta/Z\;;\;b=-\sin\theta/Z\;;\;c=\sin\theta-6\xi\gamma/Z\;;\;d=\cos\theta+6\xi\gamma/Z\;\sin\theta\;,\\
&&a_{12}=a+c/\gamma\;;\;a_{34}=d+b\gamma\;,\\
&&m^2_{r^*}=c^2 m^2_h+a^2 m^2_r\;;\;m^2_{h^*}=d^2m^2_h+b^2m^2_r\;,\\
&&r=a r^*+b h^*\;;\;h=c r^*+d h^*\;.
\end{eqnarray}

Results for the total cross-sections in the case of $c_{gp}=3.3$ with Sudakov-like suppression are
depicted in Figs.5,6 for several values of mixing parameter $\xi$ and for the vacuum expectation
value of the radion field $\Lambda_{\phi}=1$~TeV.

\section{EDD dijet production}

For the EDD production of a dijet system of mass $M_X$ in the leading order 
(see, for example~\cite{Collins}) we have:

\begin{equation}
\label{ggdijet}
|F_{gg\to X}|^2\to \frac{144\pi^2\alpha_S^2 M_X^4}{E_T^4}\;, X=gg\;,
\end{equation}

\begin{equation}
\label{qqdijet}
|F_{gg\to X}|^2\to \frac{32\pi^2\alpha_S^2 M_X^2 m_Q^2}{3E_T^4}\beta^2\;, X=Q\bar{Q}\;, \beta=\sqrt{1-4m_Q^2/M_X^2}\;,
\end{equation}
and the cross-section becomes

\begin{equation}
\label{dsigdijet}
\frac{d\sigma}{dt_1dt_2dy_Xd\kappa^{\prime}dE_T^2}\simeq\frac{|T^{Unit.}_{pp\to pjjp}|^2}{2^{13}\pi^5 s^2\kappa^{\prime}\sqrt{1-\kappa^{\prime}}}\; ,
\end{equation}
where $\kappa^{\prime}=4 E_T^2/M_X^2$, and $T^{Unit.}_{pp\to pjjp}$ is calculated as in the section 2 with
substitutions~(\ref{ggdijet}),(\ref{qqdijet}).
 
 Note that all the results of this article are given for the value of the constant $c_{gp}=3.3$, which 
is obtained from the upper bounds for the exclusive di-jet production at TeVatron 
energies~\cite{CDF}. Cross-sections and numerical estimations 
for $c_{gp}$ at different transverse energy cuts are the following:

\begin{eqnarray}
\label{dijetres}
E_T>&\!\!7\;GeV\;,& \sigma<3.7\; nb\;,\; c_{gp}<3.3\\
E_T>&\!\!10\;GeV\;,& \sigma<0.97\pm 0.065\; (stat.)\pm 0.272 (sys.)\; nb\;,\; c_{gp}<3.4\nonumber\\
E_T>&\!\!25\;GeV\;,& \sigma<34\pm 5\; (stat.)\pm 10 (sys.)\; pb\;,\; c_{gp}<4.2\;.\nonumber
\end{eqnarray}
The lowest value is close to the result, obtained by fitting the HERA data on elastic $J/\Psi$ production. It can serve as
the indication of model applicability.

From the analogous calculations for LHC with cuts~(\ref{tlimits}),(\ref{xlimits}) we have:

\begin{eqnarray}
E_T>10\; GeV,&& \sigma(pp\to p+jet+jet+p)\simeq 7\; nb\\
E_T>25\; GeV,&& \sigma(pp\to p+jet+jet+p)\simeq 150\; pb\nonumber\\
E_T>50\; GeV,&& \sigma(pp\to p+jet+jet+p)\simeq 8\; pb.\nonumber
\end{eqnarray}


\section{Conclusions}

 We see from the results that there is a real possibility to use advantages of the the EDDE 
for  investigations at LHC. Accuracy of the mass measurements could be improved by applying the missing 
mass method~\cite{Rostovtsev}.

 The low value of the exclusive Higgs boson production cross-section obtained in this paper is mainly due to
the Sudakov suppression factor~(\ref{sudakov}), the full validity of which is not obvious, because
the confinement effects can strongly modify the "real gluon emission". It is interesting that in 
spite of different models and quite different ways of account of 
absorbtive effects in our paper and in Ref.~\cite{Khoze}, the final results appeared to be quite close.

 Certainly, the cross-sections may be several times larger due to still not 
very well known non-perturbative factors.

In the case of heavy quarkonium and dijet EDD production
cross-sections are much larger than for Higgs boson production, and some other important 
investigations like measurements of the azimuthal angle dependence 
and the diffractive pattern of the interaction could be done.
 
\section*{Aknowledgements}

 We are grateful to A. De Roeck, A. Prokudin, A. Rostovtsev, N.E. Tyurin, A. Sobol, S. Slabospitsky and participants of
BLOIS2003 workshop and several CMS meetings for helpful discussions. We are also indebted
to N.V. Krasnikov and V.A. Matveev, who draw our attention to the radion phenomenology. This 
work is supported by the Russian Foundation for Basic Research, grant no. 02-02-16355




\newpage
\section*{Figure captions}

\begin{list}{Fig.}{}

\item 1: The process $p+p\to p+X+p$. Absorbtion in the initial and final pp-channels is not shown.
\item 2: The full unitarization of the process $p+p\to p+X+p$.
\item 3: t-distribution $d\sigma/dt/\sigma_{tot}$ of the process $p+p\to p+X+p$ for masses of the system X equal to 100 and 500 GeV.
\item 4: $\xi$-distribution $d\sigma/d\xi/\sigma_{tot}$ of the process $p+p\to p+X+p$ for $M_X=100$~GeV.
\item 5: The total cross-section (in fb) of the process $p+p\to p+h*+p$ versus Higgs boson($h^*$) mass for $c_{gp}=3.3$ 
with Sudakov-like suppression at LHC. Parameters of RS1 model are shown.
\item 6: The total cross-section (in fb) of the process $p+p\to p+r*+p$ versus Radion($r^*$) mass for $c_{gp}=3.3$ 
with Sudakov-like suppression at LHC. Parameters of RS1 model are shown.

\end{list}


\newpage

\begin{figure}[hb]
\label{unitar}
\vskip 4cm
\hskip  1cm \vbox to 14cm {\hbox to 16cm{\epsfxsize=16cm\epsfysize=14cm\epsffile{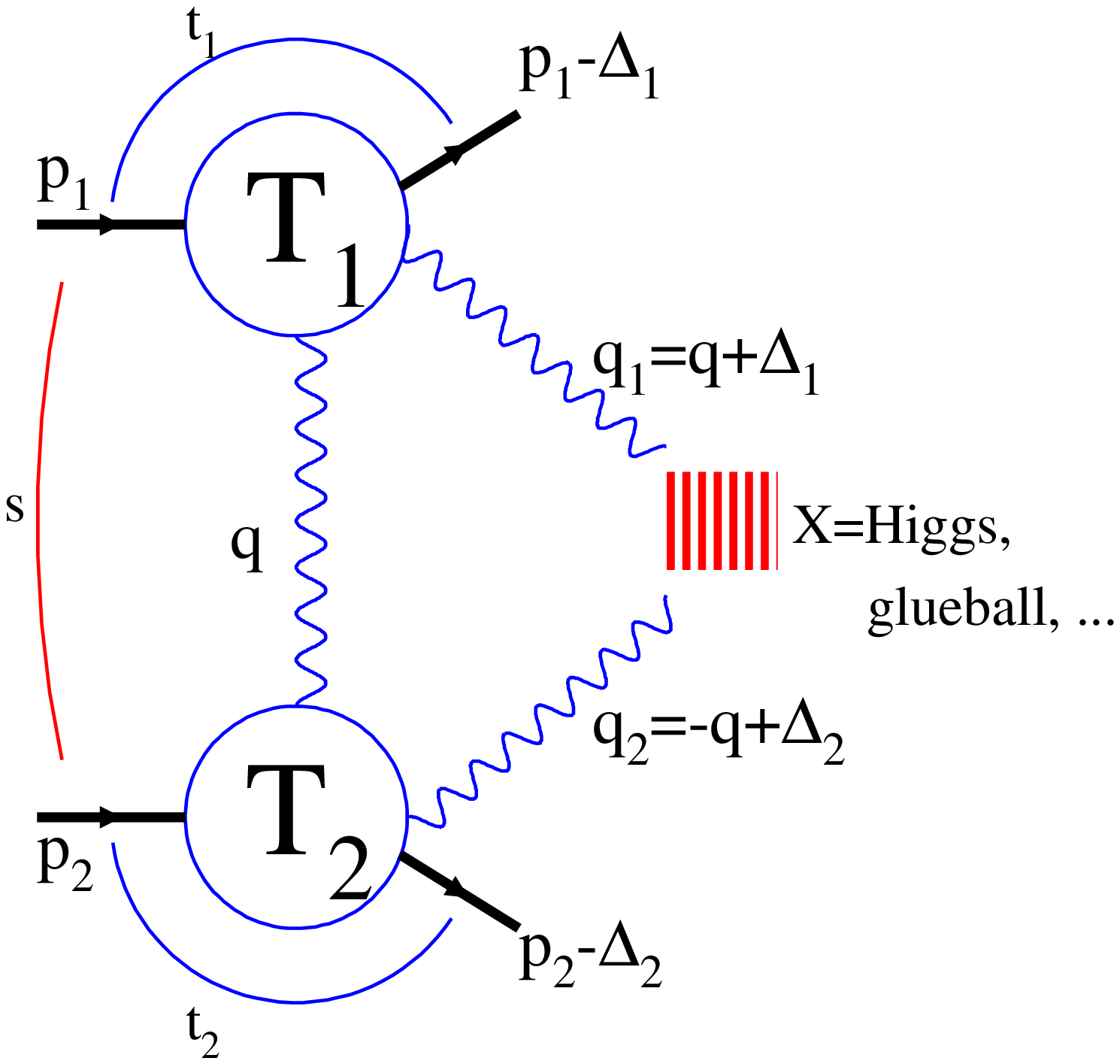}}}
\hskip 1cm
\caption{}
\end{figure}

\newpage

\begin{figure}[hb]
\label{pp_pXp}
\vskip 1.5cm
\vbox to 15cm {\hbox to 15cm{\epsfxsize=15cm\epsfysize=15cm\epsffile{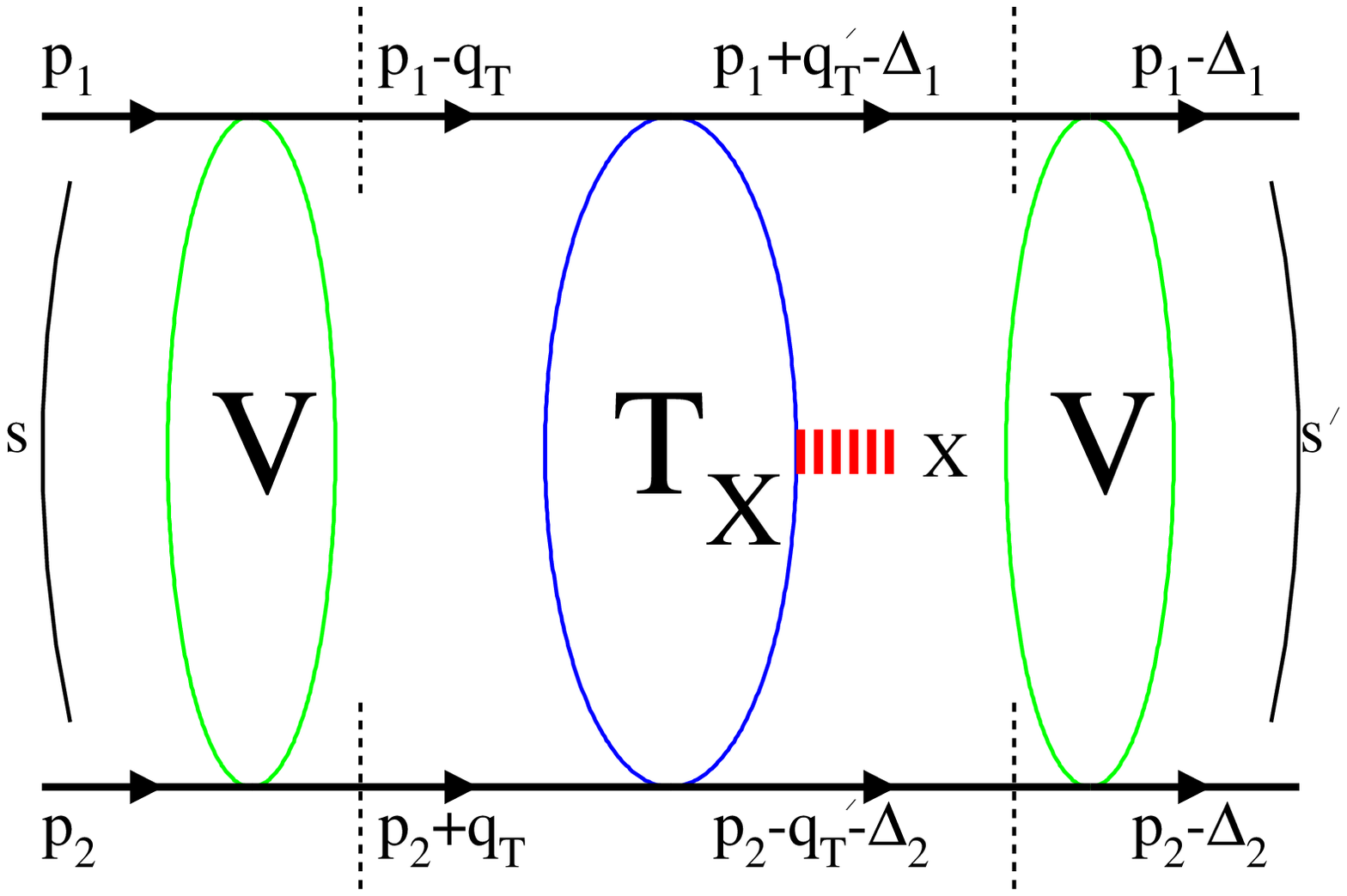}}}
\hskip 1cm
\caption{}
\end{figure}

\newpage

\begin{figure}[hb]
\label{dsigtHiggs}
\vskip 1.5cm
\vbox to 17cm {\hbox to 17cm{\epsfxsize=17cm\epsfysize=17cm\epsffile{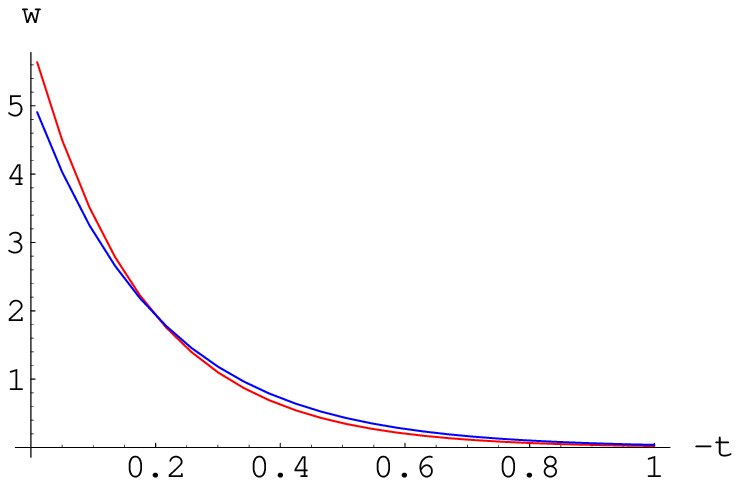}}}
\hskip 1cm
\caption{}
\end{figure}

\newpage

\begin{figure}[hb]
\label{dsigxHiggs}
\vskip 1.5cm
\vbox to 17cm {\hbox to 17cm{\epsfxsize=17cm\epsfysize=17cm\epsffile{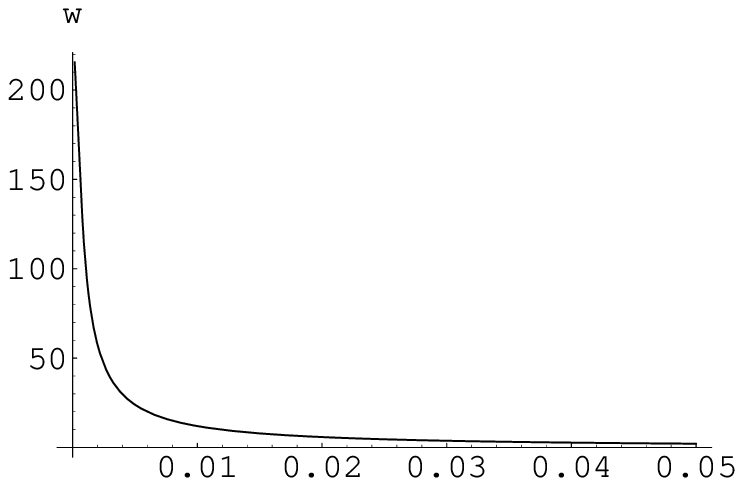}}}
\hskip 1cm
\caption{}
\end{figure}

\newpage

\begin{figure}[hb]
\label{sigtotLHC0}
\vskip 1.5cm
\vbox to 17cm {\hbox to 17cm{\epsfxsize=17cm\epsfysize=17cm\epsffile{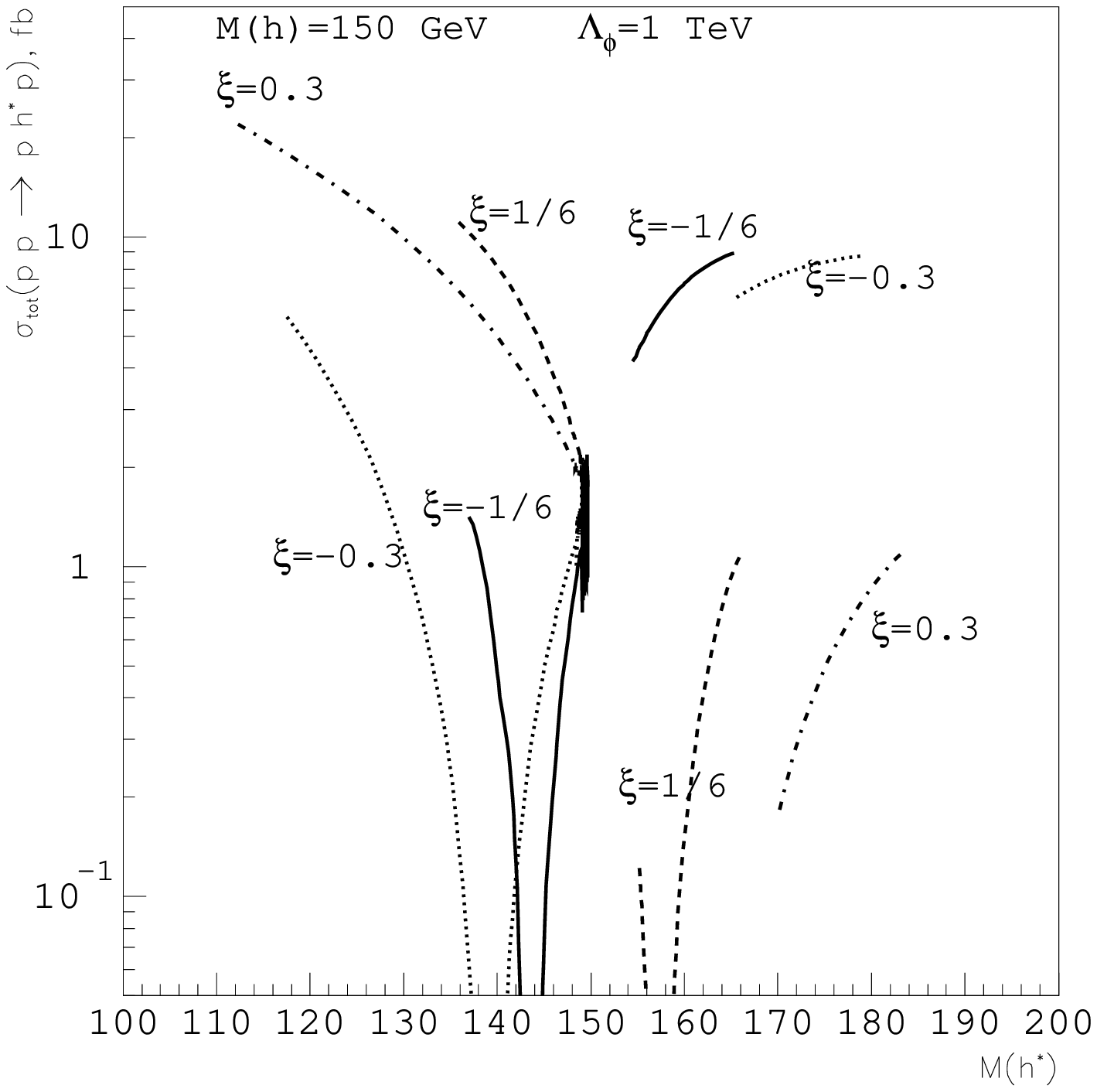}}}
\hskip 1cm
\caption{}
\end{figure}

\newpage

\begin{figure}[hb]
\label{sigtotLHC0su}
\vskip 1.5cm
\vbox to 17cm {\hbox to 17cm{\epsfxsize=17cm\epsfysize=17cm\epsffile{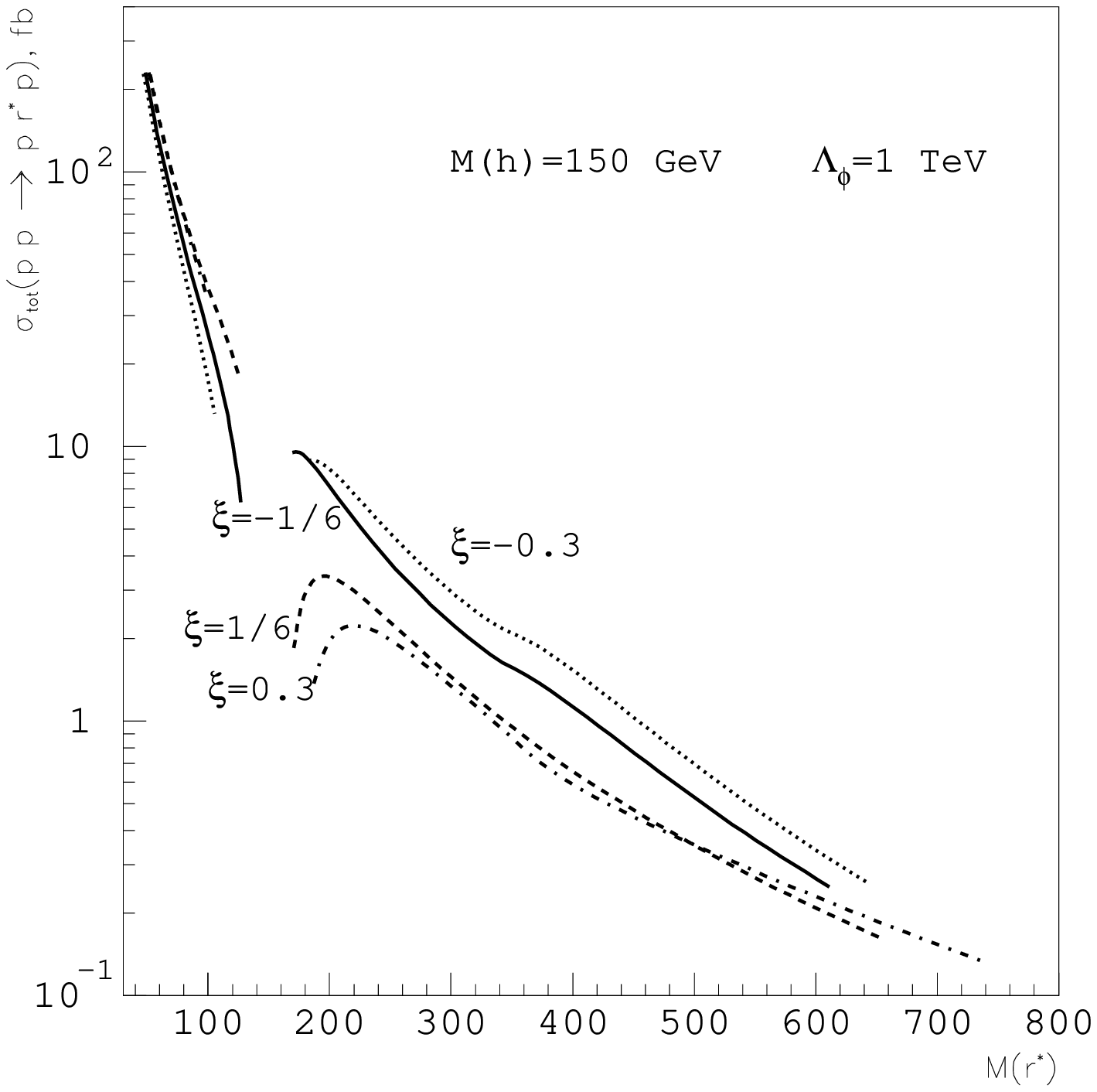}}}
\hskip 1cm
\caption{}
\end{figure}

\end{document}